\documentclass[aps,prc,superscriptaddress,reprint]{revtex4-1}
\usepackage{graphicx}
\usepackage{multirow}
\usepackage{color}

\usepackage{amssymb}
\usepackage{bbm}
\usepackage{amsmath}
\usepackage{color}
\usepackage{lineno}
\usepackage{enumitem}
\usepackage{scalerel,stackengine}
\usepackage{xcolor}
\usepackage{comment}
\usepackage[normalem]{ulem}

\def\nuc#1#2{\relax\ifmmode{}^{#1}{\protect\text{#2}}\else${}^{#1}$#2\fi}

\newcommand{\be}{\begin{eqnarray}}
\newcommand{\ee}{\end{eqnarray}}

\newcommand{\bwt}{\begin{widetext}}
\newcommand{\ewt}{\end{widetext}}

\stackMath
\newcommand\reallywidehat[1]{%
\savestack{\tmpbox}{\stretchto{%
  \scaleto{%
    \scalerel*[\widthof{\ensuremath{#1}}]{\kern-.6pt\bigwedge\kern-.6pt}%
    {\rule[-\textheight/2]{1ex}{\textheight}}
  }{\textheight}%
}{0.5ex}}%
\stackon[1pt]{#1}{\tmpbox}%
}

\begin{document}


\title{Continuum Effects and the Trojan Horse Mechanism in Halo Nuclei-Induced Reactions: Implications for Heavy Isotope Synthesis}


\author{Jin Lei}
\email[]{jinl@tongji.edu.cn}

\affiliation{School of Physics Science and Engineering, Tongji University, Shanghai 200092, China.}




\begin{abstract}
Nonelastic breakup (NEB) reactions induced by the halo nucleus $^{11}$Be on $^{64}$Zn at 28.7 MeV are investigated using the Ichimura-Austern-Vincent (IAV) model combined with the Continuum Discretized Coupled Channels (CDCC) method. NEB cross sections calculated with full CDCC wave functions (including continuum states), ground-state-only CDCC wave functions, and single-channel calculations are compared. The results indicate that continuum effects are negligible and that NEB cross sections are dominated by the ground-state contribution. This validates the use of simpler models like the distorted wave Born approximation for such reactions. Additionally, by varying the binding energy in a toy model, the feasibility of using halo nuclei in the Trojan Horse Method (THM) for synthesizing heavy isotopes is explored. It is demonstrated that THM significantly enhances sub-barrier fusion cross sections due to the weak binding of halo nuclei, offering a promising approach for the synthesis of new elements.
\end{abstract}


\date{\today}%
\maketitle

\section{Introduction}\label{sec:intro}
The study of halo nuclei has significantly advanced our understanding of nuclear forces and nucleon distribution within atomic nuclei. These nuclei are characterized by a spatially extended radius, with valence nucleons, typically neutrons, occupying regions that are classically forbidden due to their weak binding energies. This remarkable feature leads to pronounced coupling between bound states and the continuum, necessitating a detailed examination of continuum effects on halo-induced reactions. Continuum effects are especially significant because they account for breakup processes wherein the halo nucleus dissociates into its core and valence nucleons during scattering interactions. Such coupling critically influences reaction dynamics near the Coulomb barrier, affecting elastic scattering, fusion cross sections, and the overall interpretation of experimental data involving halo nuclei. This provides deeper insights into the interplay between nuclear structure and reactions in weakly bound systems~\cite{Pesudo17, Yang2022, Guimar20, Aguilera11, Pietro10, Cubero12, JP13, Hongo22, Yang21, Kubota20, Schmitt12, CANTO20151}.

In addition to the fundamental interest in continuum effects, halo nuclei offer promising pathways for synthesizing heavy and super-heavy isotopes through mechanisms like the Trojan Horse Method (THM) and incomplete fusion processes. The THM leverages the unique structure of halo nuclei to bypass the Coulomb barrier, effectively increasing reaction cross sections at energies below the barrier. In this process, an indirect approach is employed where a three-body reaction simulates the two-body reaction of interest, allowing the core of the halo nucleus to fuse with the target nucleus without experiencing the full repulsive force of the Coulomb barrier. This can lead to the formation of compound nuclei at relatively low excitation energies, below the fission barrier, thereby enhancing the probability of producing heavy isotopes.

Moreover, the known superheavy nuclei produced by heavy-ion fusion reactions occupy a narrow region of the nuclear chart close to the proton drip line; that is, these isotopes are all proton-rich systems~\cite{Nazarewicz2018}. The main challenge in synthesizing neutron-rich superheavy isotopes lies in the extremely small cross sections for their formation, even when using stable beams with higher intensities compared to neutron-rich radioactive beams. This low probability makes the synthesis of neutron-rich superheavy isotopes extremely difficult. Here, halo nuclei offer a new avenue to address this problem. Their neutron-rich cores, under the Trojan Horse condition, can be more easily incorporated into the compound nucleus, potentially leading to the production of neutron-rich superheavy isotopes.

The potential of halo nuclei in synthesis applications is further highlighted by incomplete fusion (ICF) mechanisms, where only a fragment of the projectile fuses with the target nucleus. In the breakup-fusion (BF) process, considered a component of ICF, the halo nucleus first breaks up due to its weak binding energy, and then the core fuses with the target. This sequence allows the core to bypass much of the Coulomb barrier, facilitating fusion at sub-barrier energies. By utilizing halo nuclei and methods like the THM, researchers can explore new reaction pathways that may increase the formation probability of heavy and super-heavy isotopes. These isotopes are otherwise challenging to produce due to the hindrance caused by the Coulomb barrier and the high fission probability at elevated excitation energies.

In previous work where I was a co-author~\cite{Jin19}, we investigated the BF mechanism within the ICF process using a three-body version of the Ichimura-Austern-Vincent (IAV) model~\cite{Austern87, IAV85}. We applied the Continuum Discretized Coupled Channels (CDCC) method combined with the IAV model (CDCC-IAV) to discern whether ICF proceeds predominantly through a two-step process involving the continuum (as in BF) or through a one-step mechanism involving the ground state of the halo nucleus. Our findings indicated a mixed nature for reactions induced by deuterons and $^6$Li, with the one-step process being dominant~\cite{Jin19, Jin23}. However, the specific dynamics involving halo nuclei like $^{11}$Be require further exploration to fully understand the role of continuum effects and the potential for synthesis applications.

The present study aims to deepen the understanding of continuum dynamics in reactions involving $^{11}$Be and to assess the feasibility of using halo nuclei in the THM for synthesizing heavy isotopes. By comparing nonelastic breakup (NEB) cross sections, specifically $^{64}$Zn($^{11}$Be, $^{10}$Be$X$) and $^{64}$Zn($^{11}$Be, $n$$X$), and incorporating transfer cross sections as determined by the IAV model~\cite{Udagawa89}, the impact of continuum components on the reaction mechanisms is evaluated. The results demonstrate that while there is a slight difference between the full CDCC calculations and those considering only the ground state (CDCC(gs)), the influence of the continuum on the NEB cross section and the role of the BF process are minor. Importantly, the THM shows significant potential in increasing cross sections below the Coulomb barrier between the projectile fragments (core) and the target. This suggests that employing halo nuclei with the THM could be a viable strategy for synthesizing heavy and super-heavy isotopes by enhancing fusion probabilities at sub-barrier energies.

The paper is organized as follows: In Sec.~\ref{sec:theory}, I provide a brief overview of the theoretical framework of nonelastic breakup within the IAV model. Sec.~\ref{sec:results} presents the application of the IAV model with the CDCC wave function to halo nuclei induced reactions. Finally, we conclude with a summary and discuss future research directions in Sec.~\ref{sec:sum}.

\section{Theoretical framework}\label{sec:theory}
The process under study involves a projectile labeled as $a$, which consists of a two-body structure denoted as $a = b + x$. This projectile collides with a target nucleus represented by $A$, resulting in the emission of a fragment identified as $b$. In the process, the particle $b$ acts as a \textit{spectator} and $x$ is the \textit{participant}. The process can be graphically represented as $a(=b+x) + A \to b + B^*$, where $B^*$ denotes any potential state of the $x+A$ system. The process involves both elastic scattering and nonelastic reactions between $x$ and $A$. The former is referred to as elastic breakup (EBU), while the latter is denoted by NEB. The NEB processes include inelastic scattering of $x+A$, exchange of nucleons between $x$ and $A$, fusion, and transfer to the bound state of $B$.

In the three-body model proposed by IAV~\cite{IAV85}, the cross section for the NEB inclusive process, is given by the following closed-form formula:
\begin{equation}
\left . \frac{d^2\sigma}{dE_b d\Omega_b} \right |_\mathrm{NEB} = -\frac{2}{\hbar v_{a}} \rho_b(E_b)  \langle \varphi_x (\mathbf{k}_b) | \mathrm{Im}[U_{xA}] | \varphi_x (\mathbf{k}_b) \rangle,
\end{equation}
where $\rho_b(E_b)$ is the density of states of particle $b$, $v_a$ is the velocity of the incoming particle, and $\varphi_x(\mathbf{k}_b,\mathbf{r}_{xA})$ is a relative wave function describing the motion between $x$ and $A$ when particle $b$ is scattered with momentum $\mathbf{k}_b$. The wave function $\varphi_x(\mathbf{k}_b,\mathbf{r}_{x})$ is obtained from the equation:
\begin{equation}
\label{eq:inh}
\varphi_x(\mathbf{k}_b,\mathbf{r}_{x}) =\int G_x (\mathbf{r}_{x},\mathbf{r}'_{x}) \langle \mathbf{r}'_{x}\chi_b^{(-)}(\mathbf{k}_b)| \mathcal{V}|\Psi^{3b(+)}\rangle d\mathbf{r}'_{x},
\end{equation}
where $G_x$ is the Green's function with optical potential $U_{xA}$, $\chi_b^{(-)*}(\mathbf{k}_b,\mathbf{r}_{b})$ is the distorted wave describing the relative motion between $b$ and the $B^*$ compound system (obtained with some optical potential $U_{bB}$), and $\Psi^{3b(+)}$ is the three-body scattering wave function. The post-form transition operator is given by $\mathcal{V} \equiv V_{bx} + U_{bA} - U_{bB}$, where $V_{bx}$ is the potential that binds the projectile, and $U_{bA}$ is the optical potential describing the relative scattering between $b$ and $A$. In the asymptotic limit, the wave function $\varphi_x(\mathbf{k}_b,\mathbf{r}_x)$ takes the form
\begin{equation}
\varphi_x(\mathbf{k}_b,\mathbf{r}_x) \underset{r_x \to \infty}{\longrightarrow} f(\hat{k}_b,\hat{r}_x) \frac{e^{ik_xr_x}}{r_x},
\end{equation}
where the scattering amplitude $f(\hat{k}_b,\hat{r}_x)$ can be used to compute EBU as discussed in Ref.~\cite{Jin15}.

In the three-body model, the three-body wave function $\Psi^{3b(+)}$ can be approximated as the Distorted Wave Born Approximation (DWBA) wave function $\Psi^{\mathrm{DWBA}(+)} = \chi_a^{(+)} \phi_a$, where $\chi_a^{(+)}$ characterizes the elastic scattering of $a+A$ and $\phi_a$ represents the bound state of the projectile. Although the original IAV model was proposed in the DWBA form~\cite{IAV85}, Austern et al.~\cite{Austern87} extended the model by expanding the CDCC function in relation to $b+x$ states, including continuum components,
\begin{align}
\label{eq:cdccwf}
\Psi^\mathrm{CDCC(+)} (\mathbf{r}_a,\mathbf{r}_{bx})   
&= \sum_{i}\phi_a^{i}(\mathbf{r}_{bx})\chi_a^{i(+)}(\mathbf{r}_a)   \nonumber \\
& + \sum_{n}^{N} \phi_a^{n}(k_n,\mathbf{r}_{bx})\chi_a^{n(+)}(K_n,\mathbf{r}_a),
\end{align}
where $i$ denotes the bound ground and excited states, and $n$ is a discrete index labeling the discretized continuum states.  

In conclusion, the three-body model proposed by IAV provides a framework to study the inclusive process involving the projectile $a$, target nucleus $A$, and emitted fragment $b$. The NEB cross section for this process can be calculated using the closed-form formula, and the three-body wave function can be approximated using the DWBA or expanded CDCC wave functions. 

In this study, I will use three different types of $\Psi^{3b(+)}$ as outlined in Eq.~(\ref{eq:inh}). These include:

1. The full IAV-CDCC wave function, which incorporates both the ground state and continuum couplings of $^{11}\text{Be}$, denoted as $\Psi^\mathrm{CDCC(+)}$ and described in Eq.~(\ref{eq:cdccwf}).

2. The CDCC solution focusing solely on the ground state component, $\Psi^\mathrm{CDCC(+)}_\mathrm{gs} = \phi_a^{0}(\mathbf{r}_{bx})\chi_a^{0(+)}(\mathbf{r}_a)$, representing the ground state component of $\Psi^\mathrm{CDCC(+)}$, denoted as IAV-CDCC(gs).

3. The single-channel (SC) solution, which considers only the ground state of the projectile without any continuum coupling. This is given by the solution of:
\begin{equation}
(E_{aA}-T_{a A})\big(\phi_{a}^0 | \Psi^\mathrm{SC(+)}\rangle = \langle\phi_{a}^0 | U_{bA}+U_{xA} |\phi_{a}^0\rangle \big(\phi_{a}^0 | \Psi^\mathrm{SC(+)}\rangle ,
\end{equation}
where $E_{aA}$ represents the relative energy between $a$ and $A$, and $T_{aA}$ is the kinetic energy operator for this pair system.

\begin{figure}[tb]
\begin{center}
 {\centering \resizebox*{0.9\columnwidth}{!}{\includegraphics{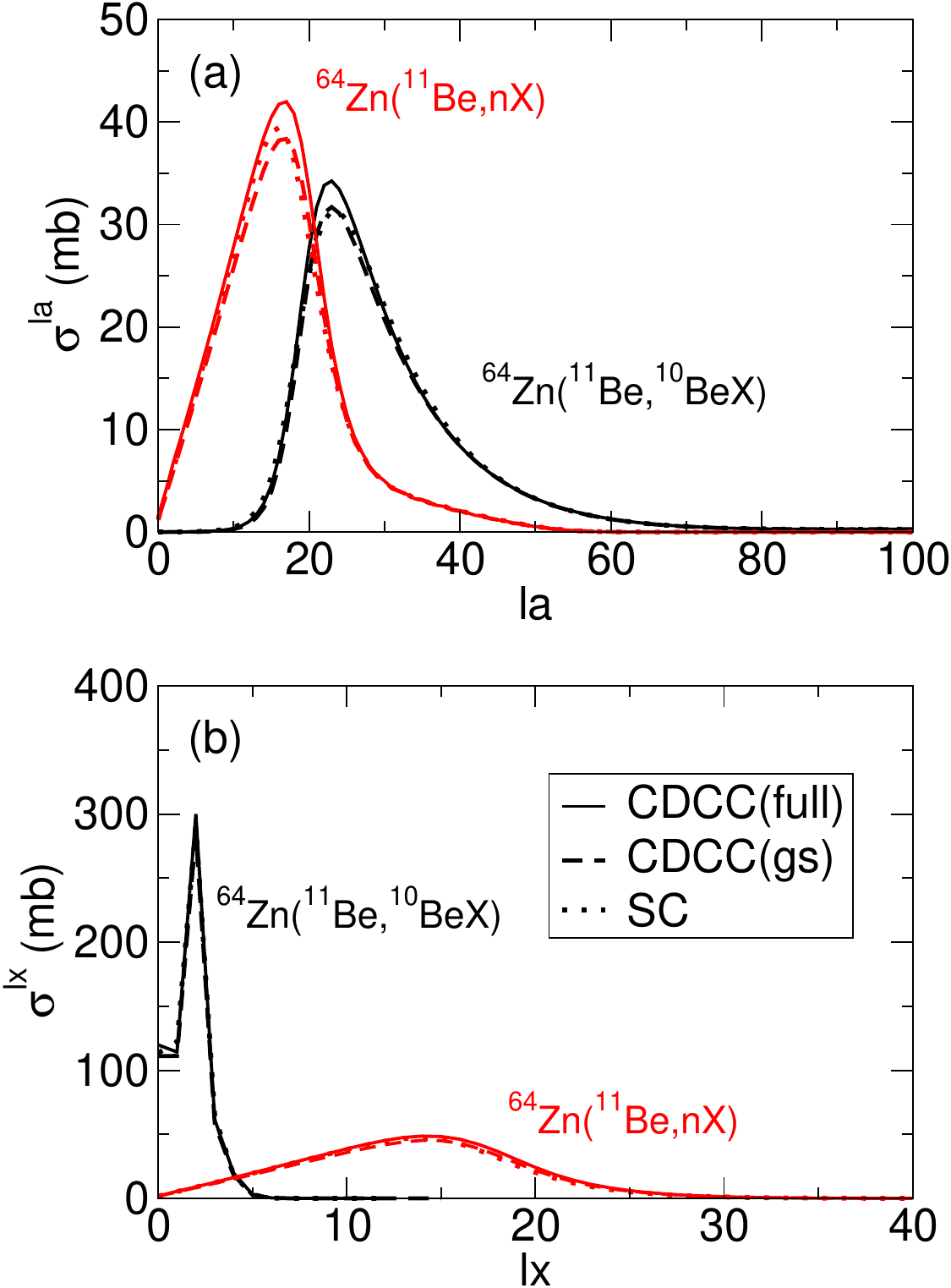}} \par}
\caption{\label{fig:compare_CDCC_SC} NEB contribution for the reaction $^{64}\text{Zn} (^{11}\text{Be}, ^{10}\text{Be}X)$, in which $^{10}\text{Be}$ acted as a spectator, and $^{64}\text{Zn} (^{11}\text{Be}, nX)$, in which $n$ plays as a spectator, at an incident energy of 28.7 MeV in the laboratory frame: (a) Cross section partial wave distribution as a function of the relative angular momentum between $^{11}\text{Be}$ and $^{64}\text{Zn}$. (b) Cross section partial wave distribution as a function of the relative angular momentum between the participant and $^{64}\text{Zn}$. See text for more details.}
\end{center}
\end{figure}

\begin{figure}[tb]
\begin{center}
 {\centering \resizebox*{0.9\columnwidth}{!}{\includegraphics{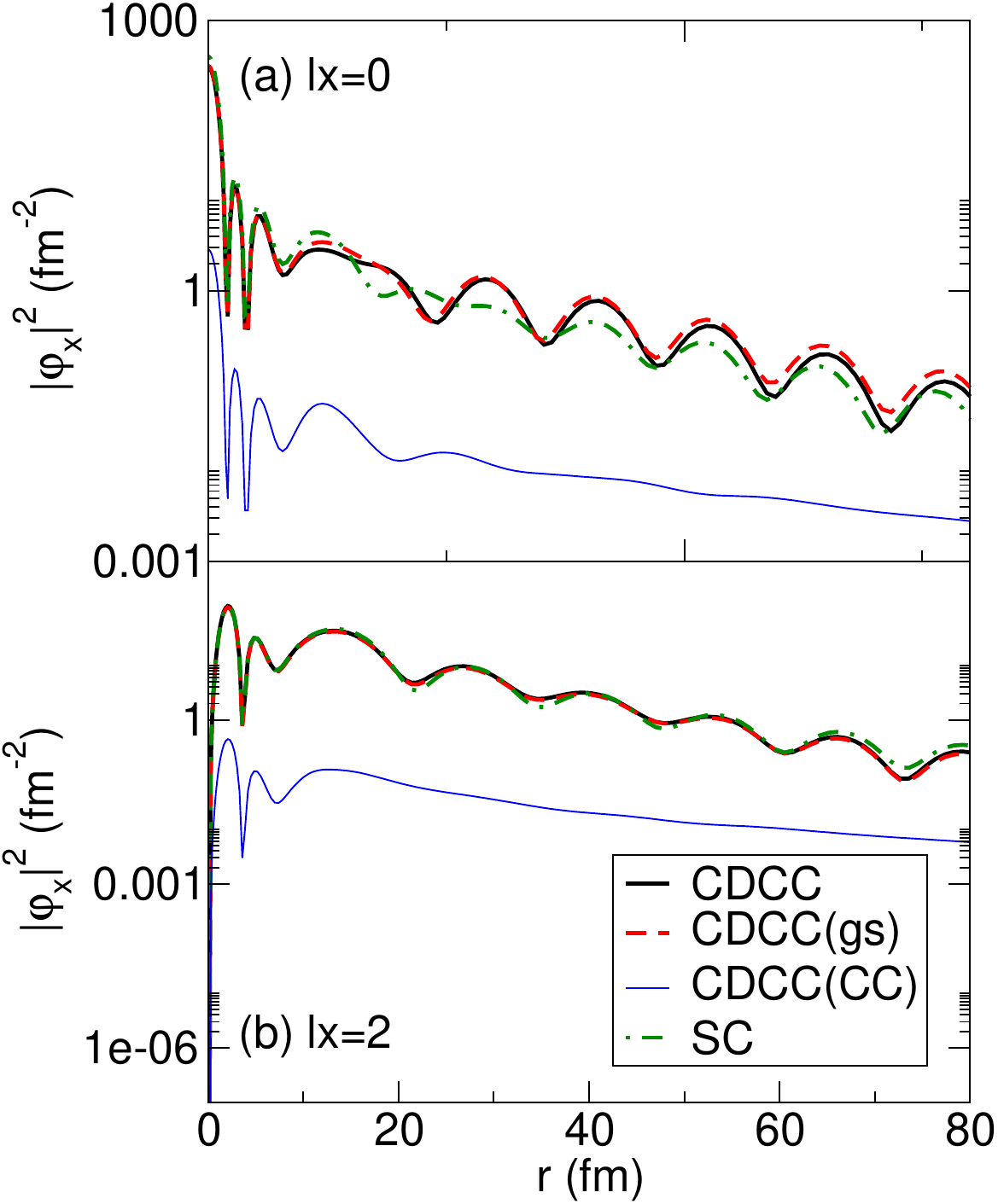}} \par}
\caption{\label{fig:compare_CDCC_SC_wf_new} Modulus squared of the radial part of the wave function between $n$ and $^{64}$Zn in the IAV model for different partial waves: (a) $\ell=0$ and (b) $\ell=2$. See text for more details.}
\end{center}
\end{figure}
\begin{figure}[tb]
\begin{center}
 {\centering \resizebox*{0.85\columnwidth}{!}{\includegraphics{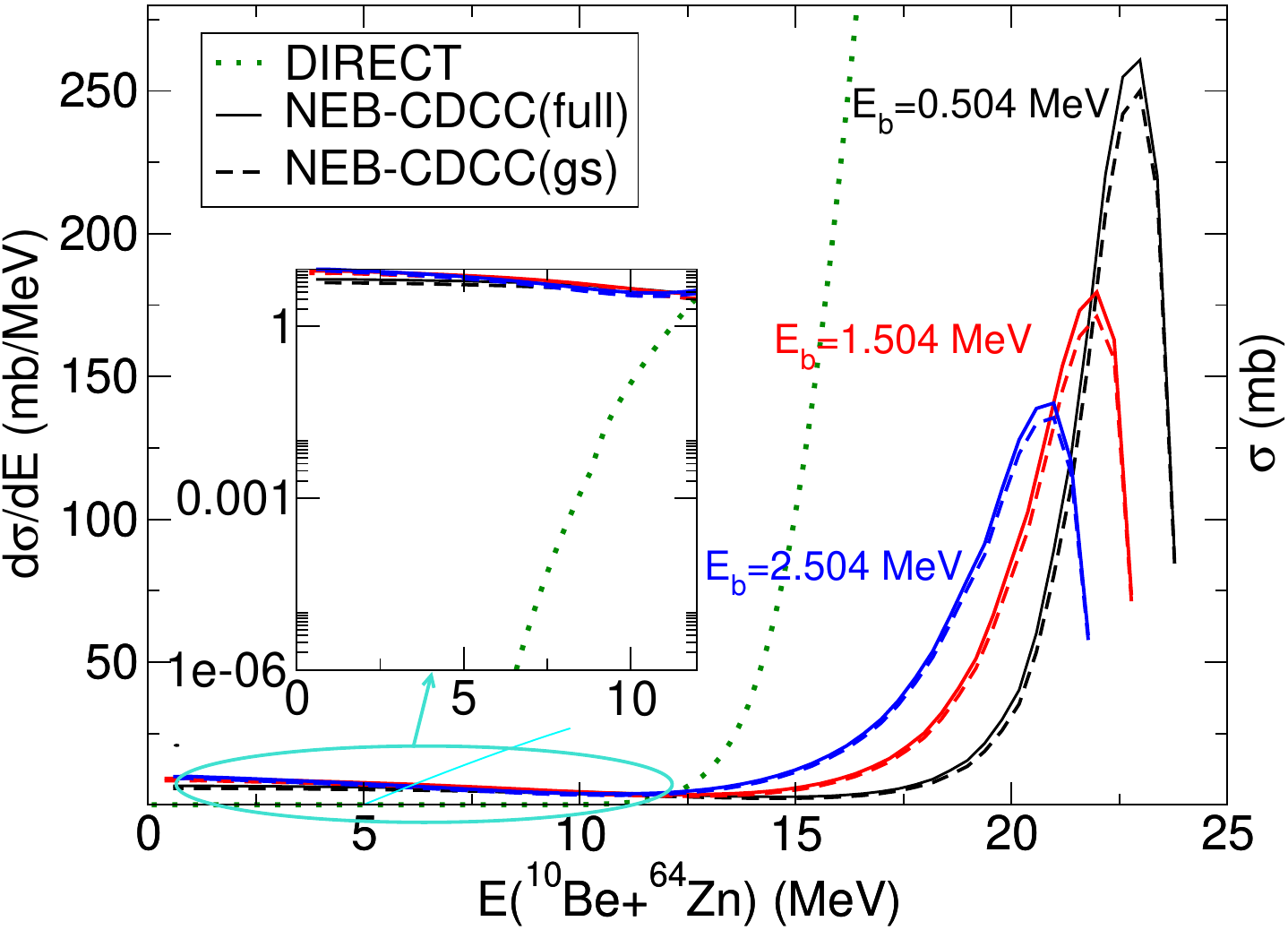}} \par}
\caption{\label{fig:compare_CDCC_CDCCgs} Toy model comparison of the NEB cross section partial wave distribution as a function of the relative momentum between $^{11}$Be and $^{64}$Zn for different binding energy of the projectile. }
\end{center}
\end{figure}

\section{Application to the $^{11}$Be induced reaction}\label{sec:results}
In this study, the breakup reaction of $^{11}\text{Be} + ^{64}\text{Zn}$ at an incident energy of 28.7 MeV in the laboratory frame is investigated. This reaction has been previously analyzed in several works~\cite{DiPietro19, Druet2012, Keeley10, DiPietro12}, which have highlighted two significant findings: the strong impact of continuum couplings on elastic scattering, and the observation of a substantial inclusive breakup cross section where $^{10}\text{Be}$ acts as a spectator and is detected experimentally. These studies integrated the EBU results from CDCC calculations and the NEB results using the DWBA version of the IAV model, showing commendable agreement with experimental data~\cite{DiPietro19}.

However, the DWBA version of the IAV model does not fully consider the continuum effects of the halo nucleus $^{11}\text{Be}$. The continuum states of $^{11}\text{Be}$ are strongly coupled to its ground state and significantly influence the elastic scattering process. To gain a deeper understanding of the role of continuum states in nuclear reactions involving halo nuclei, a detailed comparison of the NEB differential cross sections as determined by the IAV model is presented. This comparison incorporates various formulations for the three-body scattering wave function $\Psi^{3b(+)}$ as discussed in the previous section.

The interaction potentials for $^{10}$Be+$^{64}$Zn and $n+^{64}$Zn were adopted from Refs.~\cite{Pietro10} and \cite{KD02}, respectively. For the internal structure of $^{11}$Be, the single-particle neutron-core potential detailed in Ref.~\cite{Capel04} was employed. For the CDCC calculations, neutron-$^{10}$Be states up to partial waves $\ell=0-3$ are included, covering excitation energies up to 10 MeV. This range is sufficient for the current analysis, as extending the partial waves beyond this scope is necessary only for EBU calculations, which are not the focus of this study. For simplicity, any intrinsic spins of the involved particles were neglected in these calculations.

Figure~\ref{fig:compare_CDCC_SC}(a) presents the integrated NEB cross section as a function of the angular momentum between the projectile and target. The graphical representation includes solid, dashed, and dotted lines, which represent the results from IAV-CDCC, IAV-CDCC focusing solely on the ground state (IAV-CDCC(gs)), and the single-channel (SC) calculation considering only the ground state of the projectile, respectively. Notably, the results from all three calculations exhibit remarkably similar shapes and magnitudes, indicating a consistent interpretation across different modeling approaches. For interactions involving lower partial waves ($l_a < 10$), the NEB cross section is significant when the neutron ($n$) is the spectator (i.e., $^{10}\text{Be}$ is the participant). In contrast, when $^{10}\text{Be}$ is the spectator (i.e., when the neutron is the participant), the NEB cross section is nearly zero.

Conversely, Figure~\ref{fig:compare_CDCC_SC}(b) illustrates the calculated cross-section partial wave distribution as a function of the relative angular momentum between the participant and $^{64}\text{Zn}$. The overall similarity in the shape of the curves across the three methodologies reinforces the notion that the NEB processes primarily occur directly from the projectile's ground state, indicating that the continuum components have a limited influence in this specific reaction scenario. Additionally, when $^{10}\text{Be}$ acts as a participant, it exhibits a broader angular momentum distribution compared to when $n$ is the participant. This suggests that the interaction range between $^{10}\text{Be}$ and the target is significantly longer than that between $n$ and the target.

To examine the impact of differences between the ground state components and continuum state components of the CDCC wave functions on their respective NEB cross sections, Fig.~\ref{fig:compare_CDCC_SC_wf_new}(a) and (b) depict a comparison of the squared modulus of the radial part of the relative wave function between $n$ and the target. This is computed using the IAV model (Eq.~(\ref{eq:inh})) for the corresponding relative angular momenta $\ell=0$ and $\ell=2$, respectively. The thick solid lines represent the results obtained from the entire CDCC wave functions, while the dashed lines represent the results focusing on the ground state components. The thin solid lines represent the results derived by considering only the continuum components of the CDCC wave function. The dash-dotted lines represent the results from single-channel calculations that consider only the ground state of the projectile. It is observed that the full CDCC wave function yields results identical to those of the ground state component of the CDCC wave function. This indicates that, for computing the NEB cross sections in this case, the contributions from the continuum components are negligible. Consequently, this validates the use of a simple optical potential within the DWBA framework to compute the same elastic scattering observables as those obtained from the more elaborate CDCC calculations. Essentially, the DWBA approach is equivalent to using the ground-state component of the CDCC wave function, since an auxiliary potential can generate a scattering wave function equivalent to the ground-state part of the full CDCC solution.  It is noted that the continuum effects play a different role in the model proposed by Rangel \emph{et al.}~\cite{RANGEL2020135337}, the differences between the two models are discussed in the Appendix~\ref{sec:appendix}.

To explore the potential use of halo nuclei in synthesizing heavy isotopes and to discuss the effects of the projectile's binding energy, I present the NEB differential cross-section as a function of the relative energy between $^{10}\text{Be}$ and $^{64}\text{Zn}$ in a reaction where $^{10}\text{Be}$ is a participant. This is shown in Fig.~\ref{fig:compare_CDCC_CDCCgs}. To investigate the continuum effects and their variation with binding energy, I use a toy model by manually adjusting the relative binding energy between $^{10}\text{Be}$ and $n$ inside $^{11}\text{Be}$ from the actual value of $E_b = 0.504$ MeV to $E_b = 2.504$ MeV. The solid and dashed lines represent results using the full CDCC wave function (CDCC) and only the ground state part of the CDCC wave function (CDCC(gs)), respectively. Firstly, the difference between CDCC and CDCC(gs) is insensitive to changes in binding energy, indicating that the continuum state is not significantly affected by variations in binding energy in the present case. Secondly, the cross section energy distribution shows a bell-shaped curve, with a noticeable difference between CDCC and CDCC(gs) only around the peak.

To further explore the effects of the THM, I compare the direct reaction cross section of the $^{10}$Be+$^{64}$Zn system with the NEB cross section. In the analysis, the direct reaction cross section is represented by a dotted line, while the NEB cross section corresponds to the indirect method facilitated by the THM. It's important to note that these cross sections are measured in different units: mb for the direct method and mb/MeV for the NEB (indirect) method. To make a meaningful comparison between them, the NEB cross section must be integrated over energy.

Below the Coulomb barrier, the direct reaction cross section decreases exponentially because the kinetic energy is insufficient to overcome the electrostatic repulsion between the nuclei. In contrast, the indirect method, incorporating Trojan Horse effects, shows a significant enhancement—by at least three orders of magnitude—in this energy region. This substantial increase in the NEB (Nuclear Elastic Breakup) cross section enhances the probability of forming compound nuclei at low excitation energies, which is crucial for nuclear synthesis processes.

For a direct comparison, integrating the NEB cross section over the energy range from 5 to 10 MeV yields a cross section of 26 mb. In contrast, the direct method at an energy of 7 MeV results in a cross section on the order of $10^{-6}$ mb. This demonstrates a significant improvement in the reaction cross section for this region when using the THM. 

It is also important to note that the NEB cross section comprises contributions from several processes, one of which is ICF, the formation of a compound nucleus via the fusion of a participant fragment with the target. Following the approach outlined in previous studies (e.g., Ref.~\cite{Ducasse, Udagawa85}), one can partition the imaginary part of the fragment-target potential as  
\begin{equation}
    W_{xA} = W_{xA}^{\mathrm{CN}} + W_{xA}^{\mathrm{DR}},
\end{equation}
where $W_{xA}^{\mathrm{CN}}$ corresponds to the formation of the compound nucleus (ICF), and $W_{xA}^{\mathrm{DR}}$ accounts for all other direct reaction processes, such as target excitation, breakup of the fragment $x$, and nucleon exchange between $x$ and $A$. By parameterizing $W_{xA}^{\mathrm{CN}}$ (for example, using a Woods–Saxon form) and adjusting its parameters to reproduce the binary $x+A$ fusion cross section. As demonstrated in Ref.~\cite{Carlson2017}, this treatment results in a linear relation between the ICF and NEB cross section. Therefore, within this simplified picture the observed enhancement of the NEB cross section in the THM naturally implies a corresponding enhancement of compound nucleus formation at sub-Coulomb barrier energies.

This implies that the enhancement observed in the NEB cross section through the THM is not only a measure of increased nonelastic breakup probability but also indicates an enhanced probability of compound nucleus formation at sub-barrier energies.  

Furthermore, both the magnitude and the peak value of the NEB cross section depend on the binding energy of the projectile. Weakly bound nuclei are more susceptible to breakup, leading to larger NEB cross sections. This suggests that utilizing such weakly bound projectiles, particularly halo nuclei, can effectively enhance fusion probabilities at sub-barrier energies. This approach opens up promising pathways for the synthesis of new elements, including the potential use of giant halo nuclei~\cite{Meng98,Meng02,Terasaki06}, to explore uncharted territories in the periodic table through the THM.

\section{\label{sec:sum} Summary and conclusions}
This study focused on the NEB reaction of $^{11}$Be on $^{64}$Zn at an incident energy of 28.7 MeV. The results highlight two critical aspects:

Firstly, the continuum effects on the NEB process are weak. The close agreement between the full CDCC calculations and those focusing solely on the ground state (IAV-CDCC(gs) and SC) indicates that the direct breakup from the ground state is the dominant process. This consistency across different theoretical models underscores the robustness of simpler approaches, such as the DWBA, for predicting NEB differential cross sections. Consequently, detailed continuum state calculations may not be necessary for accurate predictions in similar reaction systems, simplifying future modeling efforts.

Secondly, the study underscores the potential of the THM with halo nuclei as a tool for synthesizing heavy and super-heavy isotopes. The significant enhancement of cross sections below the Coulomb barrier, driven by the unique structure of halo nuclei, suggests that THM could effectively increase fusion probabilities at sub-barrier energies. This approach opens up promising pathways for the synthesis of new elements, including the potential use of giant halo nuclei to explore uncharted territories in the periodic table.

In conclusion, while continuum effects are minimal in the NEB process, the strategic use of halo nuclei in THM offers a novel and efficient method for advancing nuclear synthesis, paving the way for the discovery of new elements.

\begin{acknowledgments}
I am grateful to Antonio Moro for his invaluable discussions. This work has been partially supported by the National Natural Science Foundation of China (Grants No. 12475132 and No. 12105204) and by the Fundamental Research Funds for the Central Universities.
\end{acknowledgments}

\bibliography{inclusive.bib}

\appendix
\section{\label{sec:appendix}Analysis of CDCC Wave Functions and Fusion Cross Section Models}
\begin{figure}[tb]
\begin{center}
 {\centering \resizebox*{0.85\columnwidth}{!}{\includegraphics{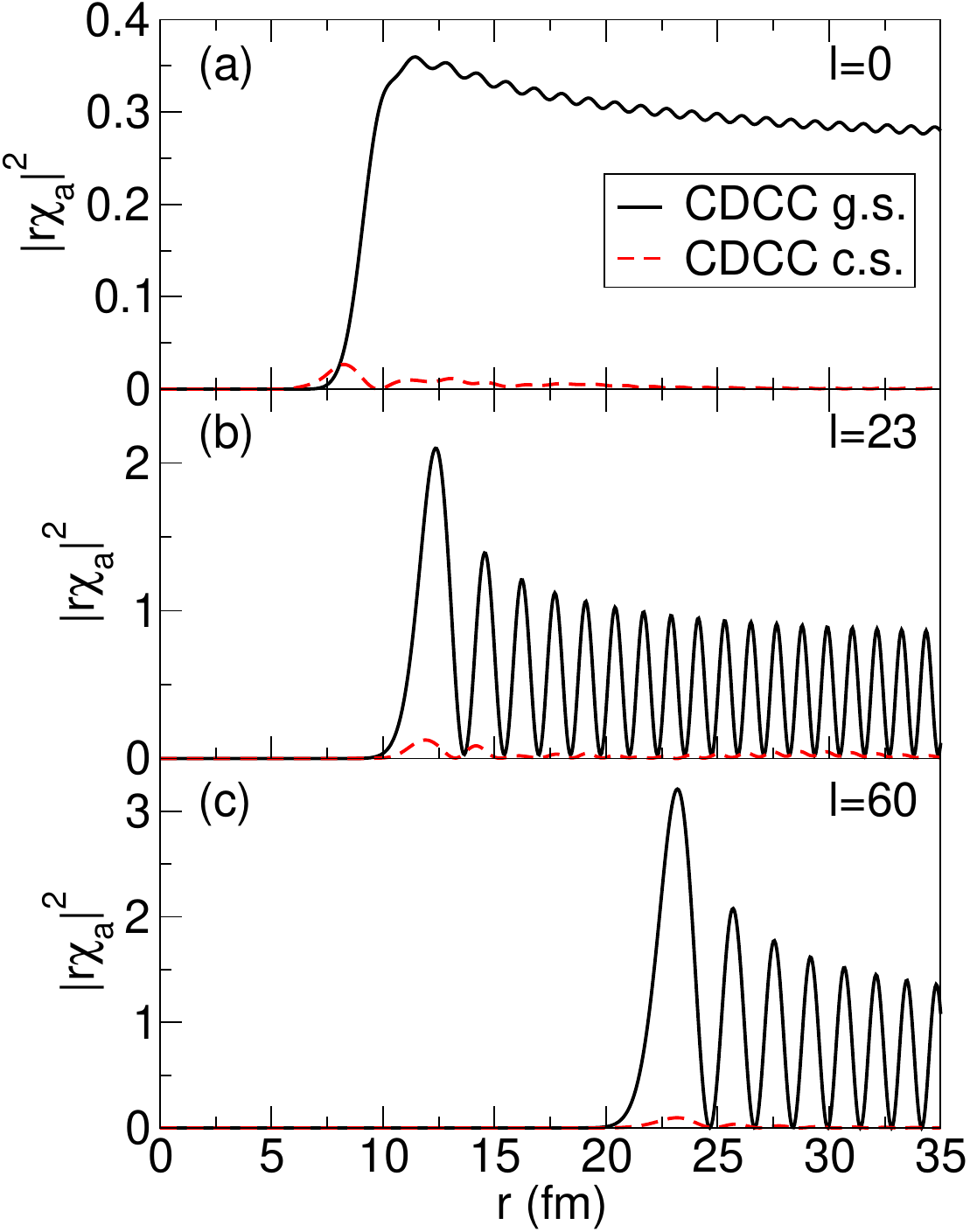}} \par}
\caption{\label{fig:com_cdcc_dwba_wf_new}The CDCC wave function in the relative coordinate between the projectile and target for different partial waves. See more details in the text. }
\end{center}
\end{figure}
In their research, Rangel \emph{et al.}~\cite{RANGEL2020135337} present an expression for the total fusion (TF) cross section that accounts for both complete and incomplete fusion processes within the framework of the CDCC method. This expression, derived in earlier studies~\cite{diaz2002effect,hagino2000role}, is given by
\begin{equation}
\label{eq:TF}
\sigma_{TF} = \frac{1}{|A|^2} \, \frac{k_a}{E_a} \sum_{\alpha,\alpha'} \langle \chi_a^{\alpha (+)} \,|\, W_{\alpha,\alpha'} \,|\, \chi_a^{\alpha' (+)} \rangle,
\end{equation}
where $A$ is the normalization constant of the scattering wave function, $k_a$ and $E_a$ are the wave number and incident energy of the projectile in the center-of-mass frame, respectively. The indices $\alpha$ and $\alpha'$ label the bound states and discretized continuum bins of the projectile's internal states, and $W_{\alpha,\alpha'}$ represents the matrix elements of the imaginary potential connecting these channels.

In their model, Rangel \emph{et al.} consider the off-diagonal matrix elements within the same subspace, either among bound states or among continuum bins, but neglect the matrix elements between different subspaces, specifically those connecting the bound and continuum states. Figure~\ref{fig:com_cdcc_dwba_wf_new} illustrates the squared modulus of the radial part of the relative wave function $\chi_a^{\alpha (+)}$ for partial waves with angular momenta $\ell = 0$, $\ell = 23$, and $\ell = 60$ in panels (a), (b), and (c), respectively. In each panel, the solid lines represent the ground-state component of the CDCC wave function, while the dashed lines depict the sum of the continuum components, $|\sum_\alpha r\, \chi_a^{\alpha (+)}(r)|^2$, which contributes significantly to the quantity used in Eq.~(\ref{eq:TF}) to compute the fusion cross section.

From these plots, it is evident that the continuum components contribute mainly in the surface region of the wave function, where the interaction between the projectile and the target is strong. In contrast, the ground-state component dominates in the asymptotic region at large distances, corresponding to elastic scattering where the projectile remains in its ground state. This behavior confirms that the full CDCC wave function appropriately describes both the elastic scattering and the coupling to the continuum states.

In the model proposed by Rangel \emph{et al.}, the short-range imaginary potential $W_{\alpha,\alpha'}$ interacts significantly with the continuum components of the CDCC wave function. Since these continuum components are substantial in the surface region where the imaginary potential is effective (as indicated by the dashed lines in Fig.~\ref{fig:com_cdcc_dwba_wf_new}), they contribute markedly to the TF cross section calculated using Eq.~(\ref{eq:TF}).

By contrast, the IAV model approaches the calculation differently. In the IAV model, the imaginary potential is applied to the subsystem consisting of one of the fragments (denoted as $x$) and the target $A$, rather than to the entire projectile-target system $a + A$ as in Rangel's model. This imaginary potential, $W_{xA}$, accounts for the absorption (or loss of flux) due to reactions between fragment $x$ and the target $A$. The IAV model evaluates the contribution from continuum processes involved in incomplete fusion, which is a part of the total fusion cross section.

The use of the imaginary potential $W_{xA}$ in the IAV model effectively suppresses the contributions from the continuum components in the region where the imaginary potential is significant. This is due to the outgoing boundary conditions imposed on the fragment-target subsystem. This suppression is evident in Fig.~\ref{fig:compare_CDCC_SC_wf_new}, where the continuum components in the IAV model are less prominent compared to those in Rangel's model.

As a result, the IAV model predicts that the incomplete fusion cross section is primarily influenced by the absorption of the fragment $x$ after the breakup, and the contribution from the continuum components is moderated by the nature of the imaginary potential acting only on the fragment-target subsystem. This leads to different predictions for the incomplete fusion cross section when compared to the total fusion cross section computed in Rangel's model, where the imaginary potential acts on the entire projectile-target system and the continuum components contribute more significantly.

This distinction emphasizes that while Rangel's model includes significant contributions from the continuum due to the direct interaction of the imaginary potential with the continuum components of the projectile-target system $a + A$, the IAV model focuses specifically on the incomplete fusion process by considering the imaginary potential on the $x + A$ subsystem and evaluating the corresponding contributions from continuum processes. The IAV model does not compute the total fusion cross section directly but provides a detailed evaluation of the incomplete fusion component arising from breakup processes.

Understanding these differences is essential for developing reliable theoretical models of fusion reactions, particularly in reactions involving weakly bound nuclei where breakup effects and incomplete fusion play a crucial role. Accurate treatment of the imaginary potential and the continuum states, and the way these are incorporated into the models, can significantly influence the calculated cross sections and improve the agreement with experimental data.

\end{document}